\documentclass[twoside]{article}
\usepackage{epsfig.sty}
\newcommand{\beq}{\begin{eqnarray}}
\newcommand{\eeq}{\end{eqnarray}}
\begin{document}

\Large {\bf $\Psi(2S)$ decay to $J/\Psi(1S)$ +2$\pi$ or $J/\Psi(1S)$ +
 $\sigma$ + 2$\pi$}
\normalsize
\vspace{5mm}

{\bf Leonard S. Kisslinger$^{1}$, Zhou Li-juan$^{2}$, Ma Wei-xing$^{3}$

1)Department of Physics, Carnegie-Mellon University, Pittsburgh, PA

\hspace{1cm} kissling@andrew.cmu.edu

2)School of Science, Guangxi University of Science and Technology,Guangxi 

\hspace{1cm} zhoulijuan05@hotmail.com

3)Institute of High Energy Physics, Chinese Academy of Sciences, Beijing

\hspace{1cm} mawx@mail.ihep.ac.cn}

\date{}

\begin{abstract}
  The BES Collaboration has measured  $\Psi(2S)$ decay to $J/\Psi \pi^+\pi^-$.
Using the mixed hybrid theory for the $\Psi(2S)$ we estimate the decay
to $J/\Psi(1S)$ + $\sigma$. Using the known $\sigma \rightarrow 2\pi$ 
coupling constant, we estimate the $\Psi(2S)$ to $J/\Psi(1S)$ + $2\pi$
decay rate and angular distribution. This is an extension of our previous 
research on $\Psi(2S)$ Decay to $J/\Psi(1S)$ +$\sigma$ + 2$\pi^o$, without 
the production of 2$\pi^o$, and with an estimate of the $\sigma \rightarrow 
\pi^+ + \pi^-$ transition.
\end{abstract}
\vspace{3mm}

 
\section{Introduction} 

  The BES Collaboration measured the $\Psi(2S) \rightarrow \pi^+ \pi^- 
J/\Psi$ decay distribution\cite{bes2000}. 
A theoretical study\cite{lsk09} determined that the $\Psi(2S)$ is a mixed
hybrid meson state, which enhances the production of the $\sigma$ two-pion
resonance. The present work is a modification of our recent research on
$\Psi(2S)$ decay to $J/\Psi(1S)$ +$\sigma$ + 2$\pi^o$\cite{kzm17}, which was
extension of previous research on  $\sigma$ production in proton-proton 
collisions\cite{lsk00,lsk05}, and the estimate of  $\sigma$ vs 2$\pi$ 
production from $\Psi(2S)\rightarrow J/\Psi(1S)+X $\cite{lsk11}. An essential
aspect of the present work is that the $\Psi(2S)$ ia a mixed hybrid charmonium
state. Using the method of QCD sum rules it was shown\cite{lsk09} that
the $\Psi(2S)$ state is approximately a 50-50 mixture of a standard charmonium
state, $|c\bar{c}(2S)>$, and a hybrid charmonium state, $|c\bar{c}g(2S)>$:
\beq
        |\Psi(2s)>&\simeq& -0.7 |c\bar{c}(2S)>+\sqrt{1-0.5}|c\bar{c}g(2S)>
 \; ,
\eeq
while the $J/\Psi(1S)$ state is essentially a standard $c \bar{c}$ charmonium
state. In Ref\cite{lsk11} it was found that the ratios of cross sections, with
the $\sigma$ a broad 2-pion resonance,
\beq
\label{lsk-krute}
         R&\equiv& \frac{\Psi(2S)\rightarrow J/\Psi(1S) + \sigma} {\Psi(2S)
\rightarrow J/\Psi(1S) + 2 \pi} \simeq 0.98 \; ,
\eeq
which is in agreement with the measurements by the BES Collaboration at
the IHEP, Beijing\cite{bes2007}.

\newpage

\section{$\Psi(2S)$ decay to $J/\Psi(1S)$ + 2$\pi$}

In this Section we derive the decay of the $\Psi(2S)$ to  $J/\Psi(1S)$ +
2$\pi$.

\subsection{The $\sigma$ mass and width}

The $\sigma$ mass and width are important for this work.

Using 14 million $\Psi(2S)$ events accumulated by BESII, the $\sigma$ mass
and width $\equiv M-i\Gamma$ were measured, Fig 2(b) in Ref.\cite{bes2007},
as shown in Figure 1
\begin{figure}[ht]
\begin{center}
\epsfig{file=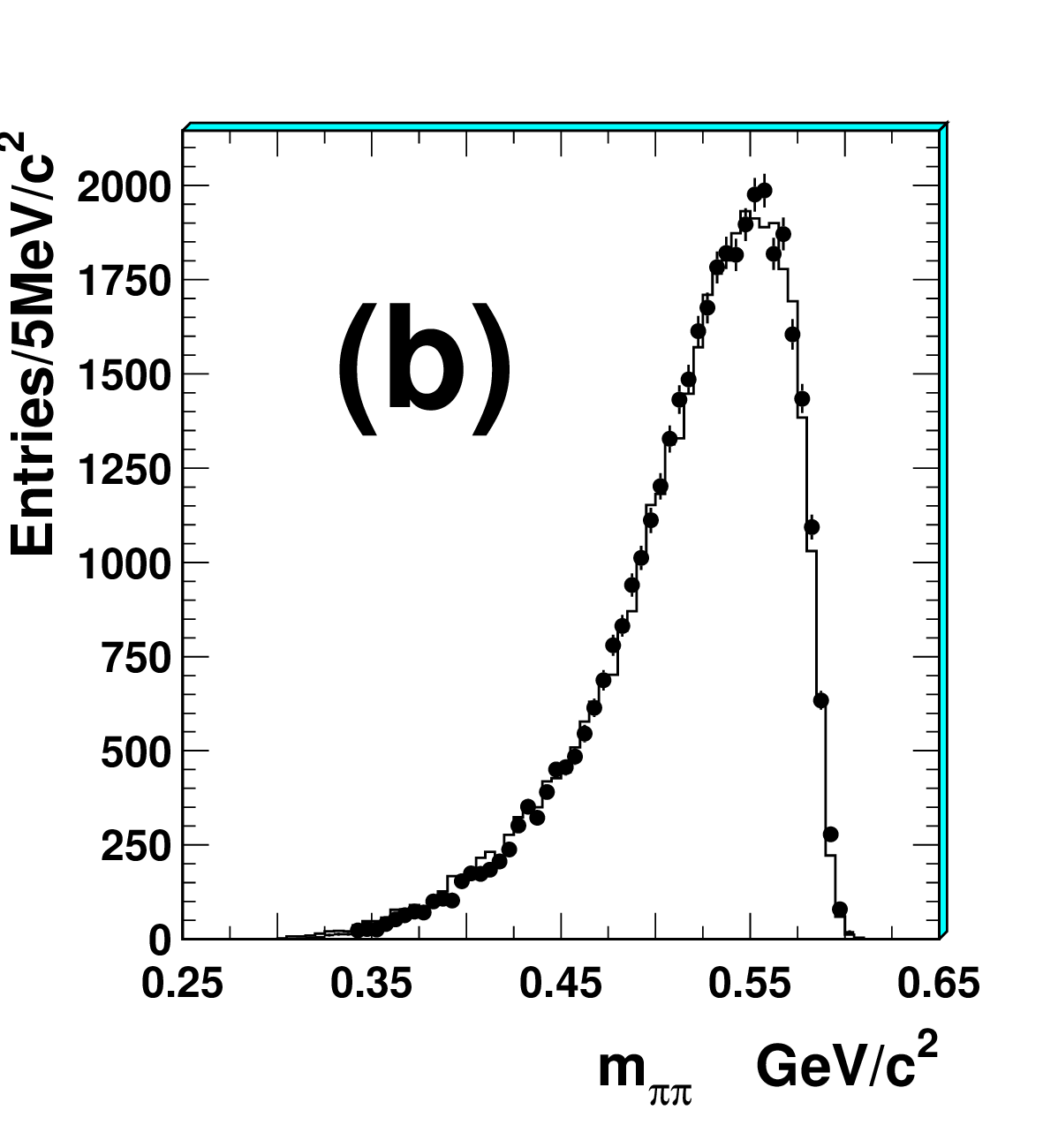,height=5cm,width=12cm}
\caption{The $\sigma$ mass and width}
\label{Figure 1}
\end{center}
\end{figure}

  From the results shown in Figure 1 in Ref.\cite{bes2007} it was estimated
  that $M-i\Gamma/2= (552^{+84}_{-106}) -i(232^{+81}_{-72})$ MeV.

  The angular distribution of the $\pi^+\pi^-$ produced by
the decay of a $\Psi(2S)$ to $J/\Psi(1S) \pi^+\pi^-$ was measured
in Ref.\cite{bes2000} , as shown in Figure 2. The $X$ in Ref.\cite{bes2000} is 
the hadron emitted by the $\Psi(2S)$ converting to $\pi^++\pi^-$, which in our 
theory is the $\sigma$, so $\theta^*_X$\cite{bes2000} is  $\theta_\sigma$.

\begin{figure}[ht]
\begin{center}
\epsfig{file=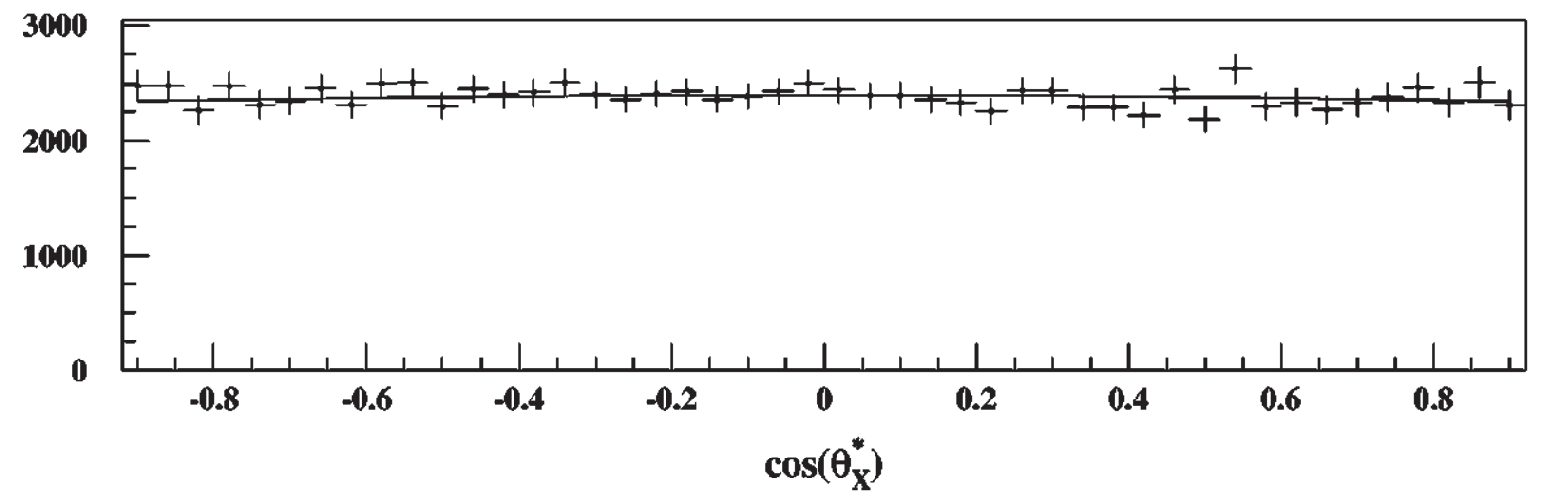,height=3cm,width=
12cm}
\caption{Angular distribution of the $\pi^+\pi^-$ produced by 
$\Psi(2S)\rightarrow J/\Psi(1S) \pi^+\pi^-$ decay from Ref.\cite{bes2000},
Fig. 5b}
\label{Figure 2}
\end{center}
\end{figure}
\vspace{1cm}
\newpage

\subsection{Estimate of $\Psi(2S)$ decay to $J/\Psi(1S)$ + 2$\pi$}

In this subection we derive the decay of the $\Psi(2S)$ to  $J/\Psi(1S)$ +
2$\pi$ using the diagram in Figure 3(b). The result can be found in 
Ref\cite{lsk11}, but we correct several typos in that article. 
\vspace{1cm}

\begin{figure}[ht]
\begin{center}
\epsfig{file=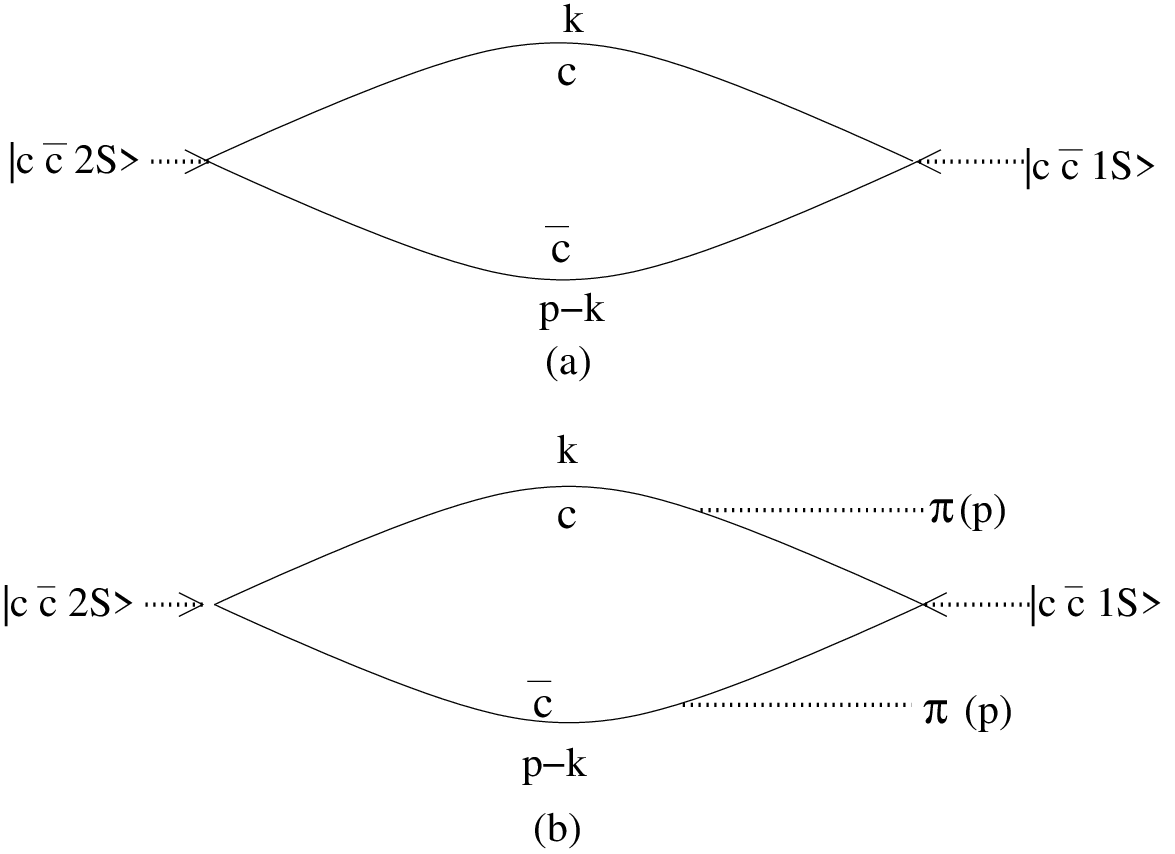,height=7cm,width=12cm}
\caption{(a)$|c\bar{c}(2S)>$= $\Psi(2S)_{normal} \rightarrow |c\bar{c}(1S)>$;
 (b)$|c\bar{c}(2S)>$= $\Psi(2S)_{normal} \rightarrow |c\bar{c}(1S)>$+$2\pi$}
\label{Figure 3}
\end{center}
\end{figure}

Our estimate of the $\Psi(2S)_{normal}$ decay to  $J/\Psi(1S)$ + 2$\pi$
first uses the correlator $\Pi_H^{\mu \nu}$ illustrated in Figure 3(a). The 
operator $J_{c\bar{c}}^\mu$ creating a $J^{PC}=1^{--}$ 
normal $c\bar{c}$ state, with $c\bar{c}$ a color singlet is
\beq
\label{Jcc}       
    J_{c\bar{c}} ^\mu&=& \sum_{a=1}^{3}\bar{c}^a \gamma^\mu c^a \; ,
\eeq
with $a$ the quark color. From this one finds
\beq
\label{PiH}
    \Pi_H^{\mu \nu}(p)&=& \sum_{a b}g^2 \int\frac{d^4k}{(4\pi)^4}
 Tr[S(k)\gamma^\mu S(p-k)\gamma^\nu ] \; ,
\eeq
where the quark propagator $S(k)=(\not{k}+M)/(k^2-M^2)$, $M$ is the
mass of a charm quark and $\not{k}=\sum_\mu k^\mu \gamma^\mu$. Since 
$Tr[S(k)\gamma^\mu S(p-k)\gamma^\nu]$ is independent of color $\sum_{a b} = 3$.
The quantity $g$, the $\Psi(nS)\rightarrow c\bar{c}$ coupling
constant\cite{lsk11} for $n=1,2$ is not needed either in the present work or in
Ref\cite{lsk11}.

\newpage
Vertices needed are illustrated in Figure 4, with G=gluon, c=charm quark,
 $\pi$=pion
\vspace{5mm}
\begin{figure}[ht]
\begin{center}
\epsfig{file=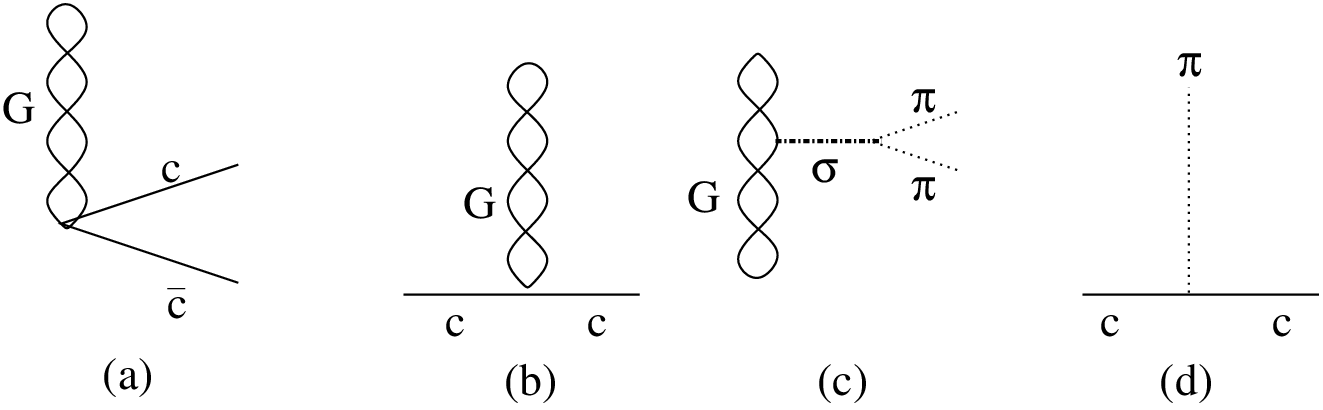,height=4.0 cm,width=12cm}
\caption{(a) gluon-charm-anticharm vertex, (b) gluon-quark coupling, 
(c) sigma-gluon coupling, (d) pion-quark coupling}
\label{Figure 4}
\end{center}
\end{figure}

The pion-quark vertex, illustrated in Figure 4d, is
\beq
\label{spik}
      S_\pi(k)&=& ig_\pi \frac{1}{\not{k}-M}=  ig_\pi \frac{\not{k}+M}{k^2-M^2}
 \; ,
\eeq
where we correct Eq(11) in Ref\cite{lsk11}, where $\frac{1}{\not{k}-M}\gamma_5
\frac{1}{\not{k}-M}$ was used rather than $\frac{1}{\not{k}-M}$.

Thus the correlator for $\Psi(2S)_{normal}$ decay to  $J/\Psi(1S)$ +2$\pi$, 
with .................$g_\pi$ the pion-quark coupling constant, from Figures(3(b(,4(d)),
is\cite{lsk11}
\beq
\label{psi-Jpsi2pi}
  \Pi_{H\pi\pi}^{\mu \nu} (p)&=& -3 g^2 g_\pi^2 \int\frac{d^4k}
{(4\pi)^4} Tr[S(k) \gamma^\mu S_\pi(k) S(p-k) S_\pi(p-k) 
\gamma^\nu] \; .
\eeq
Note if both pions illustrated in Figure 3(b) were emitted from the charm
quark, $S_\pi(k) S(p-k) S_\pi(p-k)$ in Eq(\ref{psi-Jpsi2pi}) would be
$S_\pi(k) S_\pi(p'-k)S(p+p'-k)$ and would not give $\Pi_{H\pi\pi}^{\mu \nu}(p)$.

Carrying out the trace and making use of $\frac{1}{k^2-M^2}=\int_{0}^{\infty}
d\alpha e^{\alpha(k^2-M^2)}$ one finds\cite{lsk11}
\beq
\label{Hpipi}
  \Pi_{H\pi\pi}^{\mu \nu}(p)&=& g^{\mu \nu} \frac{12}{(4\pi)^2} g_\pi^2 g^2
(2M^2-p^2/2)I_{0}(p) \nonumber \\
    &=&  g^{\mu \nu} \Pi_{H\pi\pi}^{S}(p) \; ,
 \eeq
 with $\Pi_{H\pi\pi}^{S}(p)$ the scalar component of $\Pi_{H\pi\pi}^{\mu \nu}(p)$
defined below and
\beq
\label{Iop}
  I_{0}(p)&=& \int_{0}^{1} \frac{d\alpha}{p^2(\alpha-\alpha^2)-M^2}
 \; .
 \eeq

 \subsection{Estimate of $\Psi(2S)$ decay to $J/\Psi(1S)$  + $\sigma$
 to $J/\Psi(1S)$ + $\pi^{+}$ +$\pi^{-}$}

In this subection we derive the decay of the $\Psi(2S)$ to  $J/\Psi(1S)$+
$\sigma$ to $J/\Psi(1S)$ + $\pi^{+}$ +$\pi^{-}$  using the diagram in Figure 5.
\newpage
\vspace{1cm}
\begin{figure}[ht]
\begin{center}
\epsfig{file=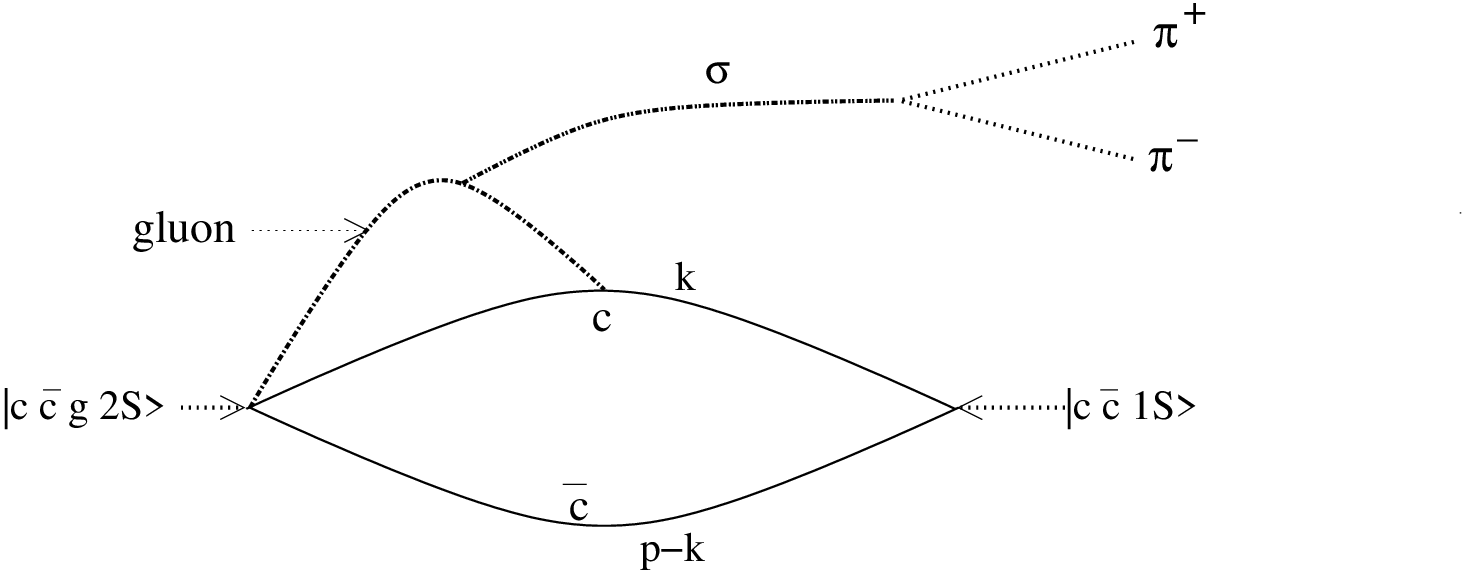,height=3cm,width=12cm}
\caption{$|c\bar{c}g(2S)>$= $\Psi(2S)_{hybrid}  \rightarrow J/\Psi(1S)+ \sigma
\rightarrow J/\Psi(1S)+ \pi^{+}+\pi^{-}$}
\label{Figure 5}
\end{center}
\end{figure}

  The estimate of the $\Psi(2S)$ decay to $J/\Psi(1S)$  + $\sigma$ 
to $J/\Psi(1S)$ + $\pi^{+} +\pi^{-}$ is made using the method of QCD
Sum Rules\cite{svz}. The correlator is obtained from $J_{HH}^\mu$, the operator 
creating the $|c\bar{c}g(2S)>$= $\Psi(2S)_{hybrid}$ state, the left vertex in 
Figure 5, which is 
\beq
\label{JHH}
          J_{HH}^\mu&=& \bar{c}^a \gamma^\nu G^{\mu \nu} c^a \;,
\eeq    
with 
\beq
\label{Gab}
       G^{\mu \nu}&=& \sum_{a=1}^{8} \frac{\lambda_a}{2}G_a^{\mu \nu} \; ,
\eeq
where $\gamma^\nu$ is the Dirac matrix, $\lambda_a$ is the QCD SU(3) generator
and $G_a^{\mu \nu}$ is the gluon field. Using the propagator of a quark of
mass M, $S(k)=1/(\not{k}+M)=(\not{k}+M)/(k^2-M^2)$, where $\not{k}=
\sum_\mu k^\mu \gamma^\mu$, the operator for the creation of the
$|c\bar{c}(1S)>$= the $J/\Psi(1S)$ state, $ J_{c\bar{c}}^\mu=
\sum_{a=1}^{3}\bar{c}^a \gamma^\mu c^a$, gluon-quark coupling =
$\frac{1}{4}S^G_{\kappa \delta}(k) G^{\kappa \delta}(0)$ with
$S^G_{\kappa \delta}(k)= [\sigma_{\kappa \delta},S(k)]_{+}$ and
$\sigma_{\kappa \delta} =i(\gamma_\kappa \gamma_\delta - g^{\kappa \delta})$.
The gluon sigma coupling is  $g_{\sigma}/M_{\sigma}<G^2>$, with 
$g_{\sigma}$ and $M_{\sigma}$ the gluon-sigma coupling constant and sigma mass
and $<G^2>$ is the gluon condensate.

  Thus the correlator for $\Psi(2S)$ decay to $J/\Psi(1S)$ +$\sigma$ to
$J/\Psi(1S) +\pi^{+}+\pi^{-}$, Figure 5, is
\beq
\label{HH2pi(+/-)-sigma}
    \Pi_{HH\sigma\pi^{+}\pi^{-}}^{\mu \nu} (p)&=& (3\frac{ g^2 g_{\sigma}}
{4M_\sigma})
g_{\sigma \pi \pi} \int\frac{d^4k} {(4\pi)^4} \nonumber \\
  && Tr[S^G_{\kappa \delta}(k)\gamma_\lambda S(p-k)\gamma^\mu]
 Tr[G^{\nu \lambda}(0) G^{\kappa \delta}(0)] \;,
\eeq
where Eq(13) in Ref.\cite{kzm17} for $\Pi_{HH\sigma\pi\pi}^{\mu \nu} (p)$ has
been modified with $g_\pi^2$ (for $2\pi^o$ production from the $c$ and
$\bar{c}$ quarks) replaced by  $g_{\sigma \pi \pi}$, the $\sigma-2\pi$ coupling 
constant.

The important new quantity, $g_{\sigma \pi \pi}$, has been shown to
be\cite{mb10}
\beq
\label{g-pi-pi}
   g_{\sigma \pi \pi}&=& \frac{m_{\sigma}^2-m_{\pi}^2}{2 f_{\pi}^2} \; ,
\eeq
where $m_{\sigma}$ is the mean mass of the $\sigma$, a $\pi-\pi$ resonance,
and $f_{\pi}$ is the pion decay constant.

  Using $m_{\sigma}= 486\pm 7$ MeV, $f_{\pi}$ =92.4 Mev, and the pion mass
$m_{\pi}\simeq 140$ Mev,  $g_{\sigma \pi \pi}$ is\cite{mmsp10}
\beq
\label{g-pi-pi-number}
   g_{\sigma \pi \pi}&\simeq& 12.8 \; .
\eeq
Note that the $\sigma\equiv f_0(500)$ mass and decay (mainly 2$\pi$) is given
in Ref\cite{ppb16}, and possible difficulties with the $f_0(500)$ resonance is
discussed in Ref\cite{jp16}.
 \newpage

 Using
\beq
\label{SGtr}
     S^G_{\kappa \delta}(k)Tr[G^{\mu \lambda}(0) G^{\kappa \delta}(0)]
&=& \frac{2i<G^2>}{96 (k^2-M^2)} 2(k_\lambda \gamma_\mu-k_\mu \gamma_\lambda)
\eeq
one finds
\beq 
\label{HH2pisigma}
    \Pi_{HH\sigma\pi\pi}^{\mu \nu} (p)&=& -\frac{3 g^2g_\sigma <G^2>g_\pi^2}
{96 M_\sigma}\int \frac{d^4k}{(2\pi)^4}\frac{Tr[(k_\lambda \gamma_\mu-k_\mu 
\gamma_\lambda)\gamma_\lambda (\not p -\not k +M)\gamma_\nu]}
{(k^2-M^2)((p-k)^2-M^2)} \nonumber \; .
\eeq

Carrying out the trace one obtains for the $HH2\pi\sigma$ correlator
\beq
\label{HH2pisigmacomplete}
    \Pi_{HH\sigma\pi\pi}^{\mu\nu}(p)&=& \frac{3 g^2g_\sigma <G^2>g_\pi^2}
{96 M_\sigma} \Pi_{\sigma\pi\pi}^{\mu\nu}(p) \; ,
\eeq
with
\beq
\label{HHpisig}
     \Pi_{\sigma\pi\pi}^{\mu \nu}(p)&=&\int \frac{d^4k}{(2\pi)^4}
\frac{4(2p^\nu+k^\nu) k^\mu} {(k^2-M^2)((p-k)^2-M^2)} \; .
\eeq

After a Borel transform $B \Pi_{HH\sigma\pi^{+}\pi^{-}}^{\mu \nu}(p)=
\Pi_{HH\sigma\pi^{+}\pi^{-}}^{\mu \nu}(M_B)$, with $M_B$ the Borel mass, finds
\beq
\label{HHpisigMB}
\Pi_{HH\sigma\pi^{+}\pi^{-}}^{\mu \nu}(M_B) &=& [\frac{12}{(4\pi)^2} 
g_{\sigma \pi \pi} g^2] \frac{g_\sigma <G^2>\pi^2}{96 M_\sigma/2}e^{-x}
(3K_0(x)-4K_1(x)+K_2(x)) \; ,
\eeq
with $x=2M^2/M_B^2$. This is the same as $B \Pi_{HH\sigma\pi\pi}^{S}(p)$, Eq(22)
in Ref.\cite{kzm17} with $g_{\sigma \pi \pi}$ replacing $g_\pi^2$. One finds for
$RRR$, the ratio of  $\Psi(2S)$ Decay to $J/\Psi(1S)$+ $\sigma\rightarrow 
J/\Psi(1S)+\pi^+\pi^-$ to $\Psi(2S)$ Decay to $J/\Psi(1S)$ + 
2$\pi^o$\cite{kzm17}

\beq
\label{RRR}
   RRR &=& -\frac{\pi^2 g_\sigma <G^2>g_{\sigma \pi \pi} (3K_0(x)-4 K_1(x)/4 
+K_2(x))}{96 g_\pi^2 M_\sigma M^2  N (K_0(x)-K_1(x))} \; ,
\eeq
with the relative $H,HH$ normalization $N\simeq 0.091$ GeV$^2$.

 Using\cite{lsk11} $K_0(x), K_1(x), K_2(x)$= 1.54, 3.75, 31.53, $g_\sigma/M_\sigma
 \simeq 1.0$, $ <G^2> = 0.476 GeV^4$, $g_{\sigma \pi \pi} \simeq 12.8$, and since
$\pi$-quark coupling is about the same for all quarks, $ g_\pi^2/4\pi 
\simeq 0.7$\cite{lein86} finds
\beq
\label{RRRvaslue}
   RRR &\simeq& 4.0
   \eeq
   
Note that $RRR$ is not directly related to experiment but is the ratio of
the Borel transforms of the correlators for  $\Psi(2S)\rightarrow J/\Psi(1S)+
\sigma \rightarrow J/\Psi(1S)+\pi^+\pi^-$ to $\Psi(2S) \rightarrow  J/\Psi(1S)
+ 2\pi^o$. Also note that we have not included the charm quark condensate
$<\bar{\psi}_c \psi_c>$ as from Ref\cite{ar12}
\beq
\label{quarkcondensate}
\frac{<\bar{\psi}_c \psi_c>}{<G^2>/M} &\simeq& \frac{0.47}{48 \pi^2} \ll 1.0
\; .
\eeq
\newpage

\section{Estimate of $\Psi(2S)$ decay to $J/\Psi(1S)$ + 2$\pi$ + 
$\sigma$}

In this Section we derive the decay of the $\Psi(2S)$ to  $J/\Psi(1S)$+
2$\pi$ + $\sigma$ using the diagram in Figure 6.

\begin{figure}[ht]
\begin{center}
\epsfig{file=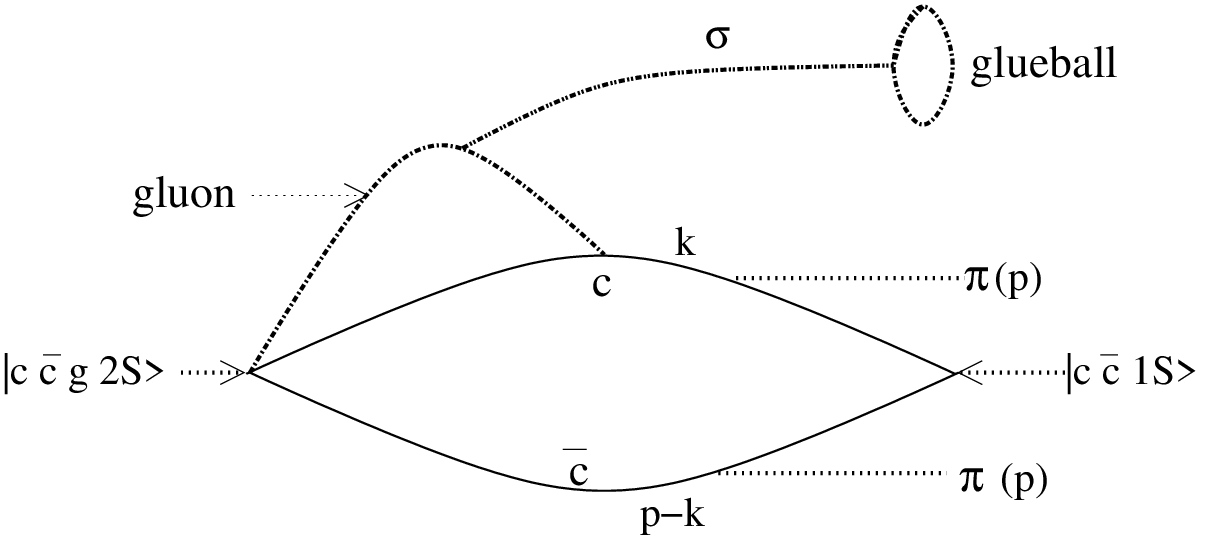,height=3cm,width=12cm}
\caption{$|c\bar{c}g(2S)>$= $\Psi(2S)_{hybrid}  \rightarrow J/\Psi(1S)+ 2\pi$ 
+ glueball via $\sigma$}
\label{Figure 6}
\end{center}
\end{figure}

Figure 6 illustrates the $c\bar{c}g$ hybrid
component of $\Psi(2S)$ decaying to $J/\Psi(1S)$ and
 two pions and a glueball via the sigma coupling to a glueball.

 Our estimate of the $\Psi(2S)_{hybrid}$ decay to $J/\Psi(1S)$+ 2$\pi$ +$\sigma$
 uses $J_{HH}^\mu$, the operator defined in Eqs(\ref{JHH},\ref{Gab}).
       
The gluon-quark coupling, Figure 4(b), is
\beq
\label{Gq}
          {\rm gluon-quark\;coupling}&=&\frac{1}{4}S^G_{\kappa \delta}(k)
G^{\kappa \delta}(0) \nonumber \\
          S^G_{\kappa \delta}(k)&=& [\sigma_{\kappa \delta},S(k)]_{+} \; ,
\eeq
with $\sigma_{\kappa \delta} =i(\gamma_\kappa \gamma_\delta - g^{\kappa \delta}$).

The gluon sigma coupling, Figure 6, is
\beq
\label{Gsigma}
      {\rm gluon-sigma\;coupling}&=& \frac{g_{\sigma}}{M_{\sigma}}<G^2> \; ,
\eeq
with $g_{\sigma}$ and $M_{\sigma}$ the gluon-sigma coupling constant and sigma mass.
$<G^2>$ is the gluon condensate.

Thus the correlator for $\Psi(2S)$ decay to $J/\Psi(1S)$+ 2$\pi$ +$\sigma$, 
Figure 5, is\cite{lsk09}  
\beq
\label{HH2pi-sigma}
    \Pi_{HH\sigma\pi\pi}^{\mu \nu} (p)&=& (3\frac{ g^2 g_{\sigma}}{4M_\sigma})
g_\pi^2 \int\frac{d^4k} {(4\pi)^4} \nonumber \\
  && Tr[S^G_{\kappa \delta}(k)\gamma_\lambda S(p-k)\gamma^\mu]
 Tr[G^{\nu \lambda}(0) G^{\kappa \delta}(0)]
\eeq

 Using
\beq
\label{SGtr}
     S^G_{\kappa \delta}(k)Tr[G^{\mu \lambda}(0) G^{\kappa \delta}(0)]
&=& \frac{2i<G^2>}{96 (k^2-M^2)} 2(k_\lambda \gamma_\mu-k_\mu \gamma_\lambda)
\eeq
one finds
\beq 
\label{HH2pisigma}
    \Pi_{HH\sigma\pi\pi}^{\mu \nu} (p)&=& -\frac{3 g^2g_\sigma <G^2>g_\pi^2}
{96 M_\sigma}\int \frac{d^4k}{(2\pi)^4}\frac{Tr[(k_\lambda \gamma_\mu-k_\mu 
\gamma_\lambda)\gamma_\lambda (\not p -\not k +M)\gamma_\nu]}
{(k^2-M^2)((p-k)^2-M^2)} \nonumber \; .
\eeq
Carrying out the trace one obtains for the $HH2\pi\sigma$ correlator
\beq
\label{HH2pisigmacomplete}
    \Pi_{HH\sigma\pi\pi}^{\mu\nu}(p)&=& \frac{3 g^2g_\sigma <G^2>g_\pi^2}
{96 M_\sigma} \Pi_{\sigma\pi\pi}^{\mu\nu}(p) \; ,
\eeq
with
\beq
\label{HHpisig}
     \Pi_{\sigma\pi\pi}^{\mu \nu}(p)&=&\int \frac{d^4k}{(2\pi)^4}
\frac{4(2p^\nu+k^\nu) p\cdot k} {(k^2-M^2)((p-k)^2-M^2)} \; .
\eeq

The Scalar components of $\Pi_{HH}^{\mu\nu}(p)=\Pi_{HH}^{S}$ and
$\Pi_{H}^{\mu\nu}(p)=\Pi_{H}^{S}$ are defined as
\beq
\label{HHpisigS}
     \Pi_{HH}^{\mu \nu}(p)&=& \frac{p^\mu p^\nu}{p^2} \Pi_{HH}^{S}(p) \\
\Pi_{H}^{\mu \nu}(p)&=& \frac{p^\mu p^\nu}{p^2} \Pi_{H}^{S}(p)
 \; ,
\eeq
where $\Pi_{H}^{\mu \nu}(p)$ is defined in Eq(\ref{PiH}) (note $g^{\mu \nu} 
p^\mu p^\nu/p^2=1$) and the correlator $\Pi_{HH}^{\mu \nu}(p)$ is obtained from 
Figure 7.
\begin{figure}[ht]
\begin{center}
\epsfig{file=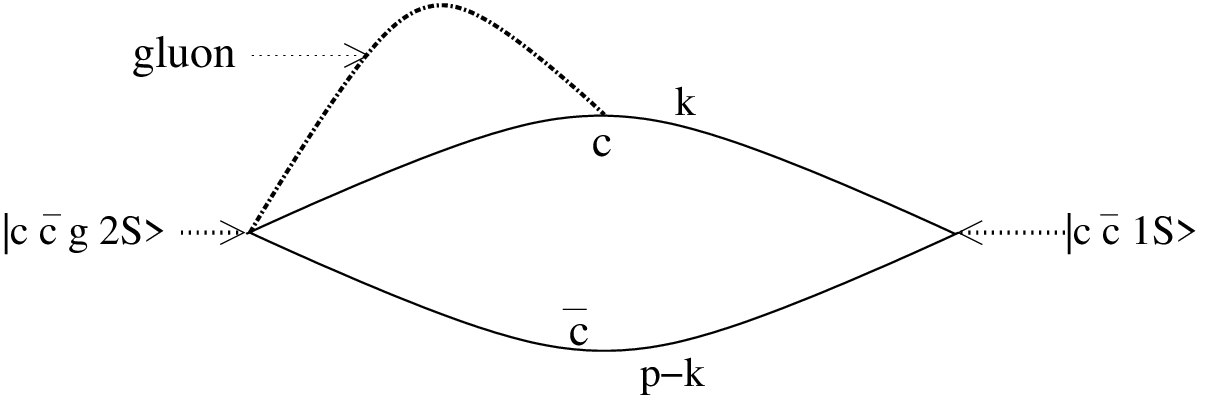,height=3cm,width=12cm}
\caption{$|c\bar{c}g(2S)>$= $\Psi(2S)_{hybrid}  \rightarrow J/\Psi(1S)+ 2\pi$ 
+ glueball via $\sigma$}
\label{Figure 7}
\end{center}
\end{figure}

   In Ref\cite{lsk09} it was shown that
\beq
\label{HHtoH}
   \Pi_{HH}^{S}(p)&\simeq& \pi^2 \Pi_{H}^{S}(p) \; .
\eeq 

Carrying out the trace and integrals in Eq(\ref{PiH}) and using
$\frac{1}{k^2-M^2}=\int_{0}^{\infty} d\alpha e^{\alpha(k^2-M^2)}$ one finds
for the scaler component\cite{lsk11}
\beq
\label{HHpisig}
     \Pi_{HH\sigma\pi\pi}^{S}(p)&=&\frac{3 g^2g_\sigma <G^2>g_\pi^2 \pi^2}
{96 M_\sigma M^2}\frac{12}{6 (4\pi)^2}(16 M^4-8M^2 p^2 + p^4) I_0(p) \; .
\eeq
\newpage

\section{Borel Transforms and Ratios of the Correlators}

To estimate the ratio of  $\Psi(2S)\rightarrow J/\Psi(1S) +\sigma+2\pi$ 
to $\Psi(2S)\rightarrow J/\Psi(1S) + 2\pi$ we use the correlators 

$\Pi^S_{H\pi\pi}(p), \Pi^S_{HH\sigma\pi\pi}(p)$, 
which are $\Pi_{H\pi\pi}^{\mu \nu}(p) ,\Pi_{HH\sigma\pi\pi}^{\mu \nu}(p)$ without 
the factor $p^\mu p^\nu/p^2$. For greater accuracy we take $B$, the Borel 
transform\cite{svz}, of $\Pi_{\pi\pi}(p),\Pi_{HH\sigma\pi\pi}(p)$. The Borel
transforms needed are (with $x=2M^2/M_B^2$, $M_B$= Borel Mass)
\beq
\label{Borel}
       B I_0(p)&=& 2 e^{-x} K_0(x) \nonumber \\
    B p^2 I_0(p)&=& 4 M^2 e^{-x} (K_0(x) + K_1(x)) \\
    B p^4 I_0(p)&=& 4 M^4 e^{-x} (3 K_0(x) +4 K_1(x)+K_2(x)) \nonumber \; ,
\eeq
where $K_n(x)$ are modified Bessel Functions. From Eqs(\ref{Hpipi},\ref{Iop},
\ref{HHpisig},\ref{Borel}) one obtains
\beq
\label{BHpipi}
   B \Pi_{H\pi\pi}^{S}(p)&=&- \frac{12}{(4\pi)^2} g_\pi^2 g^2
 e^{-x} 2M^2(K_0(x) - K_1(x)) \\
  B \Pi_{HH\sigma\pi\pi}^{S}(p)&=& [\frac{12}{(4\pi)^2} g_\pi^2 g^2]
\frac{g_\sigma <G^2>\pi^2}{96 M_\sigma/2}e^{-x}(3K_0(x)-4K_1(x)+K_2(x)) \nonumber
\; .
\eeq

Also, one must include the relative normalization, $N$, $N^2=
 \frac{\int d(M_B^2) \Pi_{H}(M_B)}{\int d(M_B^2) \Pi_{HH}(M_B)}$,
with $N\simeq 0.0123 M_B^2\simeq 4(0.0123) M^2$\cite{lsk11}, as $M_B\simeq 2M$.
From Eqs(\ref{HHpisig},\ref{Hpipi},\ref{Borel}), one finds for the ratio
of $\Psi(2S)$ Decay to $J/\Psi(1S)$+ $\sigma$ + 2$\pi$ to $\Psi(2S)$ Decay 
to $J/\Psi(1S)$ + 2$\pi$
\beq
\label{RR}
   RR &\equiv& \frac{\Pi_{HH\sigma\pi\pi}^{S}(M_B)}{N \times 
\Pi_{H\pi\pi}^{S}(M_B)} \nonumber \\
    &=& -\frac{\pi^2 g_\sigma <G^2> (3K_0(x)-4 K_1(x)/4 +K_2(x))}
{96 M_\sigma M^2  N (K_0(x)-K_1(x))} \; .
\eeq

  Using\cite{lsk11} $K_0(x), K_1(x), K_2(x)$= 1.54, 3.75, 31.53, $N \simeq$
4(.0123)$M^2$, $M\simeq 1.36 GeV$, $g_\sigma/M_\sigma \simeq 1.0$, and $ <G^2> =
0.476 GeV^4$, one
finds
\beq
\label{RRfinal}
     RR &\equiv& \frac{\Psi(2S)\rightarrow J/\Psi(1S) + \sigma +2 \pi}
{\Psi(2S)\rightarrow J/\Psi(1S) + 2 \pi}  \simeq 2.78
\eeq

Note that this is larger than $ R \equiv (\Psi(2S)\rightarrow J/\Psi(1S) + 
\sigma)/ (\Psi(2S)\rightarrow J/\Psi(1S) + 2 \pi) \simeq 0.98$ found in 
Ref{\cite{lsk11}. Note also that there are other diagrams involving gluons
in addition to the hybrid shown in Figure 2, such as a gluon arising from
$c$ or $\bar{c}$ in Figures 1 or 2, which would further increase the ratio
given in Eq(\ref{RRfinal}), and make the detection of 
$\Psi(2S)\rightarrow J/\Psi(1s)+glueball+2\pi$ more likely in future
experiments.

\newpage

\section{Angular distribution of $\Psi(2S)$ decay to $J/\Psi(1S)$  + 
$\sigma \rightarrow J/\Psi(1S)+ \pi^++\pi^-$}

To estimate the angular distribution of $\pi^+\pi^-$ and compare our prediction
to that in Ref.\cite{bes2000}, we need the momentum dependence of 
$\Pi_{HH\sigma\pi^{+}\pi^{-}}^{\mu \nu} (p)$. 

In Ref.\cite{kzm17} $\Pi_{HH\sigma\pi^{+}\pi^{-}}^{\mu \nu} (p)$ is defined as
\beq
\label{HH2pisigmacomplete}
    \Pi_{HH\sigma\pi\pi}^{\mu\nu}(p)&=& \frac{3 g^2g_\sigma <G^2>g_\pi^2}
{96 M_\sigma} \Pi_{\sigma\pi\pi}^{\mu\nu}(p) \nonumber \; ,
\eeq
with 
\beq
 \Pi_{\sigma\pi\pi}^{\mu\nu}(p)&=&\int \frac{d^4k}{(2\pi)^4}
\frac{Tr[(k_\lambda \gamma_\mu-k_\mu 
\gamma_\lambda)\gamma_\lambda (\not p -\not k +M)\gamma_\nu]}
{(k^2-M^2)((p-k)^2-M^2)} \nonumber \; .
\eeq

The trace is
\beq
  Tr[(k_\lambda \gamma_\mu-k_\mu 
\gamma_\lambda)\gamma_\lambda (\not p -\not k +M)\gamma_\nu]&=&
 4g^{\mu \nu} k\cdot(p-k) -12k^\mu p^\nu-4k^\nu p^\mu+16k^\mu k^\nu \; ,
\eeq
therefore
\beq
 \Pi_{\sigma\pi\pi}^{\mu\nu}(p)&=&\int \frac{d^4k}{(2\pi)^4}
\frac{4g^{\mu\nu}k\cdot(p-k) -12 k^\mu p^\nu-4k^\nu p^\mu+16k^\mu k^\nu}
{(k^2-M^2)((p-k)^2-M^2)} \; .
\eeq

After carrying out the $\int \frac{d^4k}{(2\pi)^4}$ integrals 
one obtains, with $I_o(p)=\int_0^1\frac{d \alpha}{(p^2(\alpha-\alpha^2)-M^2)}$,
\beq
 \Pi_{\sigma\pi\pi}^{\mu\nu}(p)&=& -g^{\mu \nu}\frac{4}{(4\pi)^2}
\{p^2(p^2/4-M^2)I_o(p)-7p^2/4] +M^2(2M^2-p^2/2)I_o(p)\} \; .
\eeq

Note that $\Pi_{\sigma\pi\pi}^{\mu\nu}(p)$ is a function of $p^2$ and is
independent of the angle $\theta_{X=\sigma}$ at which 
$\sigma \rightarrow \pi^+ \pi^-$ is emitted by $\Psi(2S)$ decaying to 
$J/\Psi(1S)$. This is consistent with the experiment\cite{bes2000} as shown 
in Figure 2.

\section{Conclusions}
  Our conclusion is that the ratio $(\Psi(2S)\rightarrow J/\Psi(1S)+
\sigma \rightarrow J/\Psi(1S)+\pi^+\pi^-)/(\Psi(2S)\rightarrow J/\Psi(1S)+
2\pi^0)$ is large enough to be measured, and might be detected in future BES 
experiments. We estimated the angular distribution of $\Psi(2S)$ decay to 
$J/\Psi(1S)$  + $\sigma \rightarrow J/\Psi(1S)+ \pi^++\pi^-$ and found a 
distribution independent of $\theta_\sigma$, the emission angle of $\pi^+\pi^-$,
which is consistent within errors of the experimental measurement\cite{bes2000}.
\vspace{3mm}

\newpage
\Large

{\bf Acknowledgements:}
\vspace{2mm}

\normalsize
{\bf Author L.S.K. acknowledges support from the P25 group at 
Los Alamos National Laboratory. Author Zhou Li-juan acknowledges the support
in part by the National Natural Science Foundation of China (11865005) and
Guangxi Natural Science Foundation (2018GXNSFAA281024) .} 
 
\end{document}